\documentstyle[sprocl,epsf]{article}

\def\be{\begin{equation}}
\def\ee{\end{equation}}
\def\bea{\begin{eqnarray}}
\def\eea{\end{eqnarray}}

\newcommand{\bef}{\begin{figure}}
\newcommand{\eef}{\end{figure}}
\newcommand{\hmp}{ h^{-1}Mpc}
\newcommand{\etal}{{\it et al.}}
\def\spose#1{\hbox to 0pt{#1\hss}}
\def\ltapprox{\mathrel{\spose{\lower 3pt\hbox{$\mathchar"218$}}
 \raise 2.0pt\hbox{$\mathchar"13C$}}}
\def\gtapprox{\mathrel{\spose{\lower 3pt\hbox{$\mathchar"218$}}
 \raise 2.0pt\hbox{$\mathchar"13E$}}}
\def\inapprox{\mathrel{\spose{\lower 3pt\hbox{$\mathchar"218$}}
 \raise 2.0pt\hbox{$\mathchar"232$}}}

\begin{document}

\title{Fractals and the galaxy distribution
\footnote{In the Proceedings of the 2nd Intas Metteing 
"Fundamental problems in classical, quantum and string gravity"}}

\author{F. Sylos Labini }

\address{ D\'ept.~de Physique Th\'eorique, Universit\'e de Gen\`eve,  
		24, Quai E. Ansermet, CH-1211 Gen\`eve, Switzerland.
		and 
		   INFM Sezione Roma1,        
		      Dip. di Fisica, Universit\'a "La Sapienza", 
		      P.le A. Moro, 2,  
        	      I-00185 Roma, Italy. }

\maketitle
\abstracts{
 There is a general agreement that galaxy  
 structures 
 exhibit fractal properties, at least up to some small scale. 
 However the presence of an eventual
 crossover towards homogenization, as well as the exact value of 
 the fractal dimension, are still a matter of debate. 
 I summarize the main points of the this discussion,
 considering also some galaxy surveys which have 
 recently appeared.
 Further I discuss the implications
 for the standard picture of the observed 
 fractal behaviour in galaxy distribution.
 In particular I consider 
 the co-existence of 
 fractal structures and the linear Hubble-law 
 within the same scales. This fact represents a challenge 
 for the standard cosmology where the linear Hubble law is a strict 
 consequence of homogeneity of the expanding universe. 
 Finally I consider the comparison of CDM-like models with 
 the data noting that the 
 simulations are not able to reproduce
 the observed properties
 of galaxy correlations.
 }

\section{Introduction} 
 
It has been  well known for
 over twenty years \cite{devac70} \cite{man77} 
\cite{pee80} that galaxy structures exhibits fractal properties 
at small scales ($\sim 10 \hmp$). The scale invariant correlated 
behavior corresponds to the existence of large scale 
structures (hereafter LSS) in the galaxy distribution. 
This evidence came from 
the analysis of the angular galaxy catalogs
and from few sparse measurements of redshifts. 
At scale larger than $\sim 10 \hmp$ 
it was reasonable, from the angular data alone,  
to assume the homogeneity of galaxy  
distribution, that is the cornerstone of the Big Bang  
model. 
 
However, with the extensive  
measurements of redshifts started in the eighties  
it was discovered that the extent of galaxy structures 
and voids is limited only by the size of the  
available samples. Such a situation has been confirmed 
by several recent 3-d galaxy catalogs. 
It is important to note that at the present time 
the investigation of the large scale distribution 
of galaxies has having  an exponential growth which 
will lead, in less than ten years, to collect  
more than one million  of galaxy redshifts. 
Such a situation, together with the use  
of the modern methods of statistical physics  
for a quantitative characterization of the distribution, 
gives rise to a rather  
different perspective on the properties 
of the large scale distribution of matter in the universe 
as well as on the theoretical methods adopted. 
Very recently a wide debate on this subject is
in progress: see the web page
 {\it http://www.phys.uniroma1.it/DOCS/PIL/pil.html}
 where all these materials have been collected.

\section{The problem of the usual perspective} 
 
The usual analysis \cite{pee80} \cite{pee93} \cite{pee98} of  
galaxy distribution identifies a correlation length 
of about $\sim 5 \hmp$. Such a length scale should  
characterize the distance at which the density fluctuations 
is of the order the average density. Pietronero \cite{pie87} 
criticized this result on the basis
that such a small characteristic length 
is actually inconsistent with the existence of 
LSS and huge voids larger  more than one order 
of magnitude in size. The problem of the standard analysis of 
galaxy correlation lies in the a priori assumption of homogeneity.
In other terms 
one usually defines an average density $\langle n \rangle$ 
in a given galaxy sample, and then one compare the density
fluctuations 
$\delta n$ to such 
a value. Such a procedure does not allow one test whether the
average density is a  
meaningful quantity. However from  the studies of 
irregular systems \cite{man77} we have 
learned that in a self similar structure there are
no characteristic values and concepts 
like the average density cannot be defined properly. 
More specifically such quantities are not  related to  the nature 
of the distribution, rather they depend on the size of the sample,
the unique 
meaningful length-scale which can be defined in such a situation. 
 
\section{Methods of analysis} 
 
Let us briefly illustrate the methods of statistical analysis 
usually used in the studies of irregular, self-similar, 
structures,  but that can be successfully used also  
for the characterization of regular systems. 
Pietronero \cite{pie87} proposed to study the
 conditional average density defined as 
\be 
\label{eq1} 
\Gamma(r) = \frac{1}{S(r)} \frac{dN}{dr} = \frac{BD}{4\pi}r^{D-3} \; . 
\ee 
Such a quantity measures the number of points $dN$ in a spherical  
shell of thickness $dr$ and volume $S(r)=4 \pi r^2 dr$, located  
at distance $r$ from an occupied point. Then 
one determines the average over all the points contained in a given
sample. Being an average quantity, $\Gamma(r)$ is a rather 
stable and robust statistical indicator. 
The last equality in eq.\ref{eq1} holds for a fractal 
with dimension $D<3$ and prefactor $B$. If the distribution 
is homogeneous ($D=3$) $\Gamma(r)$ equals 
the average density in the sample.  The power law behaviour 
of $\Gamma(r)$ implies the self-similarity of the distribution. 
The prefactor $B$ can be defined for real structures contained in finite  
samples. It gives the normalization of the amplitude of the space density. 
A very simple interpretation of such a quantity is the following: 
in a homogenous distribution (according to eq.\ref{eq1}) $B$  
is the space density (a part trivial prefactor). The average distance  
between nearest particles is known to be $\Lambda \sim B^{1/3}$. 
In the case of a fractal distribution it is possible to show \cite{slmp98} 
that the average distance between nearest neighbors is of the 
order $\Lambda \sim B^{1/D}$ (this is an intrinsic
quantity that does not depend on sample size).
 Note that while in a homogeneous 
distribution $\Lambda$ gives also a reasonable order of magnitude 
for the typical voids contained in the sample, in the fractal case 
the size of the voids scales  as a function of the size of the sample 
and it is closely related to another property 
called {\it lacunarity}\cite{man77} \cite{man97}.  
 
From eq.\ref{eq1} it follows that for a fractal distribution, 
the average density in a sample of radius $R_s$ scales as 
$R_s^{D-3}$ and hence it does not represent a  meaningful 
reference value. This simple observation shows that 
the usual statistical analysis based on concepts like  
  $\xi(r)$, the  
power spectrum and other related quantities becomes 
meaningless, unless a very well define  transition 
to homogeneity is present in the sample. 
In terms of $\Gamma(r)$ this transition should 
be shown by the break of the power law  into a 
flatter behavior with scale.  
 
It is simple to show \cite{slmp98} that all the characteristic length 
scales usually identified in the study of the LSS become  
spurious and dependent on the sample size, unless the  
density shows a constant behaviour with scale.  
Some example are: the so-called correlation length $r_0$ 
(scale at which $\xi(r_0) =1$), the turnover scale $\lambda_s$  of the  
power spectrum $P(k)$  (defined ad $dP(\lambda_s(k))/dk =0$), the  
scale at which the density fluctuations is $\delta N/N =1$.
(For a more detailed discussion see 
Sylos Labini \etal, 1998 \cite{slmp98}).

\section{Review of main results} 

A real fractal structure can be observed and defined in finite samples. 
Hence it is important to clarify the lower and the upper cut-offs  
among which its properties can be properly studied by a  
suitable correlation analysis. As already mentioned the lower cut-off 
$\Lambda$
is an intrinsic quantity of a given fractal and it gives
the order of magnitude 
of the average distance between nearest neighbors. 
The upper cut-off $R_s$ 
is defined to be the radius of the  
maximum sphere fully contained in the sample volume 
\cite{cp92} \cite{slmp98}. This definition avoids any assumption  
on the treatments of the boundary conditions of the  
sample in the correlations analysis. 
 
In  the range of scale $\Lambda < r < R_s$ we have found, in  
a complete  and extensive study of all the available redshift catalogs 
\cite{slmp98}, that galaxy distribution has well defined
fractal properties. 
In particular, by the full correlation analysis 
we have found a fractal dimension $D =2.1 \pm 0.2$ 
in the range of scales $\sim 0.5 < r < 150 \hmp$ (Fig.\ref{fig1}).
 This result is  
substantially stable in the different catalogs we have considered. 
A recent analysis of the SSRS2 and CfA2 South galaxy samples\cite{sl98}
is in complete agreement with all the other galaxy surveys previously
considered. Also the ESP survey\cite{joyce} shows
a continuation of the fractal behavior with $D\approx 2$ up
to $\sim 600 \div 800 \hmp$, although this result
is much weaker from a statistical point of view (see Joyce \etal, 
1998 for a more detailed discussion on this subject).

The prefactor $B$ is found to be
\be
B= 10 \div 15 \hmp^{-D}
\ee
and hence the lower cut-off is of the order 
$\Lambda \sim 0.5 \hmp$. This is the minimum
statistical distance beyond which the statistical
properties are well defined. However
in real samples on should consider also
a luminosity selection effect which 
can cause to change  $\Lambda$
by more than one order of magnitude, depending
on the volume limited sample considered.
\begin{figure}
\epsfxsize 10cm
\centerline{\epsfbox{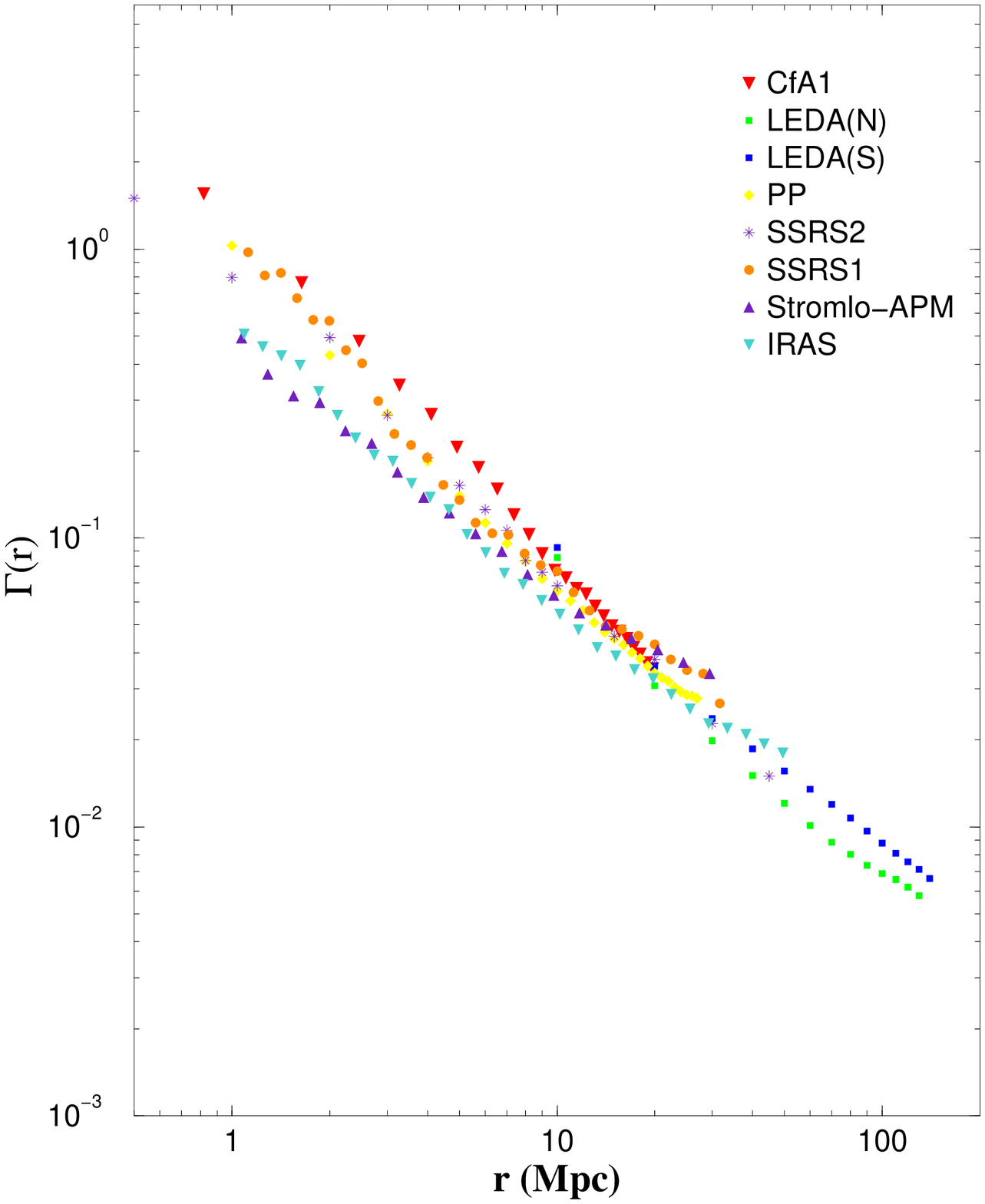}}
\caption{\label{fig1} Full correlation analysis for the various
available redshift surveys in the range of distance $0.5 \div 1050
\hmp$. The  slope is $-1$, which 
corresponds to fractal dimension $D = 2$.
(From Sylos Labini \etal, 1998). 
} 
\eef

\section{Problems of the Standard FRW scenario}
 In his classic paper, Hubble\cite{hu29} found a roughly linear 
relation between the spectral line displacement  
$z =(\lambda_{obs}-\lambda_{em})/\lambda_{em}$ 
of the line emitted by a far away galaxy $  \lambda_{em}$,
and its distance $r$.  The empirical 
Hubble Law may be written as 
\be 
\label{e5} 
cz = H_0 r \; ,  
\ee 
where $c$ is the velocity of light and $H_0$ is the Hubble 
constant.  As an observationally established relation, the Hubble 
law does not refer to any interpretation of redshift.   
Space expansion and Doppler mechanisms in falt space 
 for redshift, yields 
at first order to Eq.\ref{e5}.  If redshift is interpreted as a motion 
effect, then 
\be 
\label{e6} 
V \approx H_0 r 
\ee 
where $V$ is either space expansion velocity $v_{exp}$  
or ordinary velocity of a body moving 
in the Euclidean space.  Usually this 
velocity-distance relation is called the Hubble Law, but it is 
more correct to regard it as 
the redshift-distance relation of Eq.\ref{e5}.
This is based on the primarily measured quantities 
(redshift and distance), while velocity is inferred from redshift 
in the frame of some cosmological model.

Since its discovery, the validity of the Hubble law has been  
confirmed in an ever increasing distance  
interval where local and more remote distance indicators may be 
tied together.  Recently, several new distances have been measured  
to local galaxies using observations of Cepheid variable stars, 
thanks to the Hubble Space Telescope programmes\cite{tam96}.
Along with previous Earth-based Cepheid distances,  
methods like Supernovae Ia and Tully-Fisher have been better 
calibrated than before and confirm the linearity with good 
accuracy up to $z \approx  0.1$.  Brightest cluster galaxies 
trace the Hubble law even deeper, up to $z \approx 1$, and 
radio galaxies have provided such evidence at still larger 
redshifts\cite{san95}.

It is well known that there are small deviations $\delta V$ from 
the Hubble velocity $V_H$, connected with local mass 
concentrations such as 
the Virgo Cluster, and, possibly the Great Attractor. 
However, these perturbations are still only of the order 
$\delta V/V_H \sim 0.1$, while in the general  
field the Hubble law has been suggested to be quite smooth, 
with $\delta V$ around $50 km/s$\cite{san95}.

 Without  the actual knowledge of   matter distribution, the linearity 
and the small scatter of the observed  
Hubble law for field galaxies would make one   guess 
that the galaxies are uniformly distributed: as it was asserted 
above, this is the basis for the linear Hubble law in the standard 
cosmology.  In fact, it  
has been a common supposition that when the Hubble law was found  
in the nearby space, one finally had entered a cosmologically  
representative region of the Universe. At the same time, 
it has been clear  
that at small distances where Hubble found his  
relation, the galaxy distribution is quite inhomogeneous. 
Though, it has been believed that beyond some, not too large 
distance, the distribution should become uniform. 
 
As we have already mentioned,  
studies of the 3-dimensional galaxy universe have shown 
that de Vaucouleurs' prescient view on the matter distribution 
is valid at least in the range of scales  $\sim 0.5 \div 200 \;Mpc$
(hereafter $H_0=55 km/sec/Mpc$). 
The Hubble and de Vaucouleurs laws describe very different  
aspects of the Universe, but both have in common universality and  
observer independence.   
This makes them fundamental cosmological laws and it is important 
to investigate the consequences of their coexistence at 
similar length-scales.  In Fig.\ref{yurifig2} 
we display these laws together. 
\begin{figure}
\epsfxsize 8cm
\centerline{\epsfbox{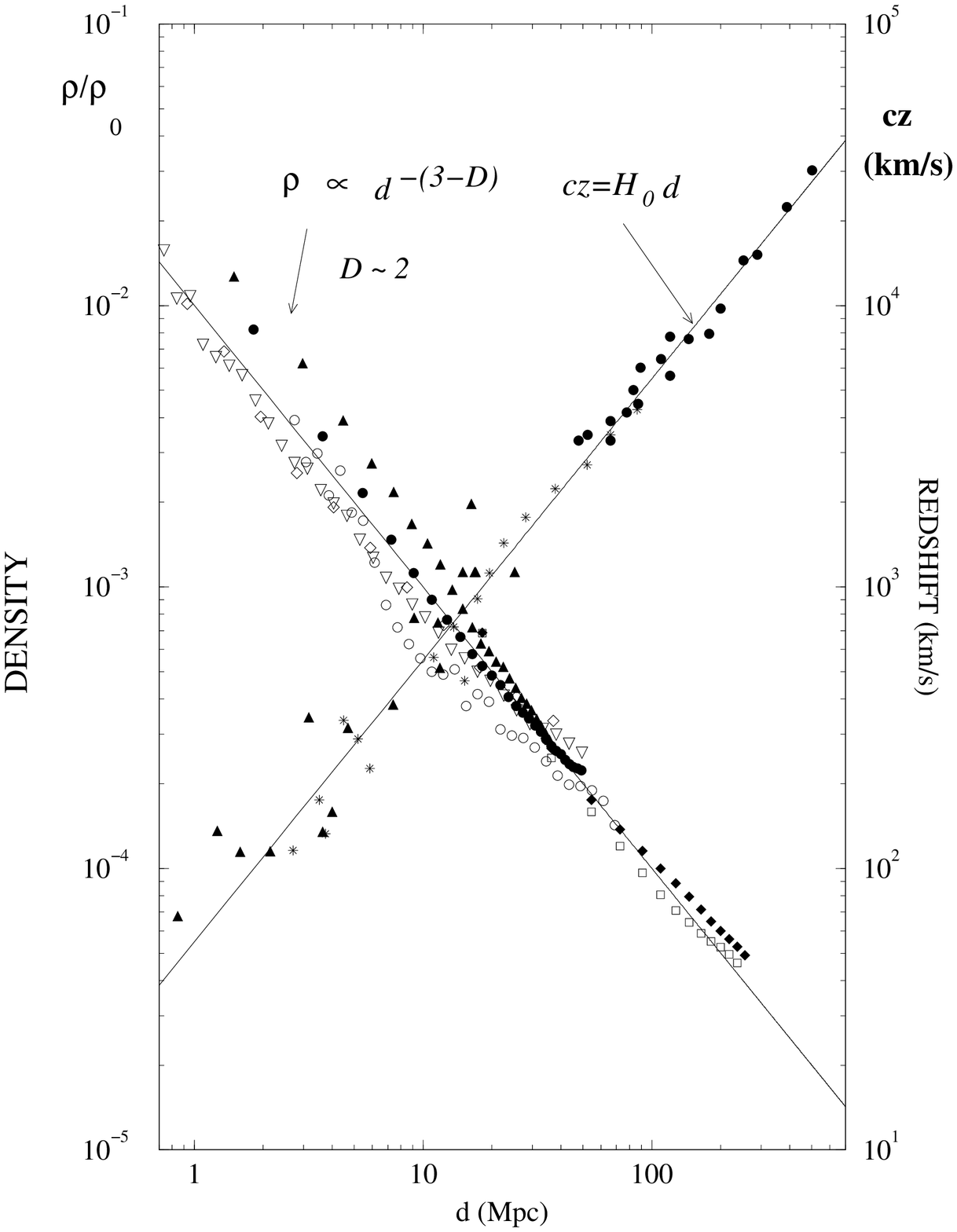}}
\caption{\label{yurifig2}  
  Hubble redshift-distance and de Vaucouleurs density-distance  
laws in the distance scales from $1$ to  
$500 Mpc$ 
 ($H_0=55 km/s/Mpc$).  
 The Hubble law (increasing from left to right) is constructed 
from: galaxies with Cepheid-distances for $cz > 0$ (triangles),  
galaxies with Tully-Fisher 
(B-magnitude) distances (stars), galaxies with SNIa-distances for $cz > 3000 
km/s$ (filled circles). 
 TF-distance points are generally averages of a few tens of galaxies 
from the "unbiased plateau" of the method of normalized distances.  
Redshift $cz$ is reduced to the Local Group center and contains 
the small correction due to the Virgo infall velocity field.  The solid 
line  corresponds to the Hubble law with  $H_o = 55 km/s/Mpc$. 
The de Vaucouleurs law (decreasing from left to right) in the normalized 
form is constructed from  
the computation of the conditional average density in the following 
 redshift surveys: 
CfA1 (crosses), Perseus-Pisces, LCRS (filled diamonds), 
ESP (triangles left) 
and LEDA (Sylos Labini \etal, 1998). 
The normalization between the different densities 
takes into account the different magnitude limited of the various  
redshift surveys. 
The dotted line  corresponds to the de Vaucouleurs law with 
 correlation exponent $\gamma = 1$, i.e. $D=2$.
(From Baryshev \etal, 1998) } 
\eef 
A representative Hubble law has been taken from Fig.4 of Teerikorpi 
(1997)\cite{tee97},
based on Cepheid distances to local galaxies, Tully-Fisher 
distances from the KLUN programme, and Supernovae Ia distances. 
The behavior of the conditional density 
(De Vaucouleurs law) presented in Fig.\ref{yurifig2} has been 
taken from  
Sylos Labini \etal (1998).

The puzzling conclusion from Fig.\ref{yurifig2} is that the strictly linear  
redshift-distance relation is observed deep inside the  
fractal structure.  (Note that in the analysis of galaxy redshift surveys
one uses the Hubble law for the distance determinations 
as an experimental fact, i.e. any assumption has been used).
This empirical fact presents a profound 
challenge to the standard model where the  
homogeneity is the basic explanation of the Hubble law, and "the  
connection between homogeneity and Hubble's law was the first success of  
the expanding world model" \cite{pee91}.  
This also reminds us 
the natural reaction  of several authors: 
"In fact, we would not expect any neat relation of 
proportionality between velocity and distance  
[for such close galaxies]" \cite{wei77}.

However, contrary to the expectations, modern data show a good linear  
Hubble law even for nearby galaxies.  How unexpected this actually is, 
can be expressed quantitatively for the standard model and is briefly 
discussed 
below (for a more detailed discussion see Baryshev \etal, 1998).

According to the standard Big Bang model the universe obeys 
 Einstein's 
Cosmological Principle: it is homogeneous, isotropic 
and expanding \cite{wei72}\cite{pee93}.
Homogeneity of matter distribution is the  
central hypothesis 
of the standard cosmology because it allows one to introduce the space  
of uniform curvature in the form of the Robertson-Walker line element $R(t)$.
This line element leads immediately to a linear  
relation between velocity and proper distance.
Indeed, consider a comoving body at a fixed coordinate distance 
from a comoving observer.  At cosmic $t$,
let $l= R(t) y$ be the  proper distance 
from the observer.
The expansion velocity $v_{exp} = dl/dt$, 
defined as the rate of change of the proper distance $l$,
is 
\be 
\label{e2} 
v_{exp} = H l = c \cdot  \frac{l}{l_{H}}
\ee 
where $H = \dot{R}/R$ is the Hubble constant and $l_H = c/H$ 
is the Hubble  
distance.  In this way, the linear velocity-distance relation of  
Eq.\ref{e2}  
is an exact formula for all Friedmann models and a  rigorous consequence 
of spatial homogeneity. 
In particular, for $l > l_H$, the expansion velocity $v_{exp} > c$.   
Such an apparent violation of  
special relativity is consistent with general relativity 
\cite{her93}.

In the expanding space the wavelength of an emitted photon is  
progressively stretched, so that the observed redshift $z$ is  
given by Lemaitre's redshift law 
\be 
\label{e3} 
z = \frac{R(t_{obs})}{R(t_{em})} - 1 
\ee 
which is a consequence of the radial null-geodesic of the FRW
line element. For $z \ll 1$ Eq.\ref{e3} yields  
$z \approx dR/R \approx H_0 dt \approx l/l_H$,  
and from Eq.\ref{e2} one gets the approximate velocity-redshift  
relation that is valid for small redshifts 
\be 
\label{e4} 
v_{exp} \approx cz \; . 
\ee 
We note that
the  
expansion velocity-redshift relation differs from the  
relativistic Doppler 
effect.  So, the space expansion redshift 
mechanism in the standard model is quite distinct from the usual  
Doppler mechanism.  We stress this points, because in the literature 
these two redshift mechanisms are often confused. 
 In the context of the standard  
cosmology, it has been natural to interpret the Hubble Law  
as a reflection of  Eq.\ref{e4}  
and to regard the coefficient of  
proportionality $H_0$ in Eq.\ref{e5}  
as the present value of the theoretical Hubble constant $H$  
from Eq.\ref{e2}.

We consider now the case of an expanding universe,
where an average density is well defined and has a constant value $\rho_0$.
In such a case, by neglecting relativistic 
effects  and the terms depending on pressure, 
according to the linear approximation, there is a velocity
deflection $\delta V$ from the unperturbed Hubble flow $V_H=H_o r$
in the scale where the density perturbation is $\delta \rho$.
In the case of zero cosmological constant
and {\it spherical mass distribution}, this deflection has grown during
the Hubble time to the present value which is (Eq.20.55 from
Peebles, 1993):
\be
\label{pek1}
\frac {\delta V} {V_H} = \frac { 1}{ 3} {\Omega_0}^{0.6}
\frac {\delta \rho}   {\rho_0}
\ee
where $\Omega_0 = \rho_0/\rho_{crit}$ is the density parameter of
the Friedmann model.
This approximation holds in the limit $\delta \rho/\rho_0\ll 1$.

Let us consider a two-component model for the density distribution in
a Friedmann universe.  
First, there is the component which exhibits fractal behavior up to  a maximum 
scale, and which we call
$\lambda_0$.  At larger scales   
this component is homogeneous with an average density 
$\rho_{lum}$.
The second  component is  dark matter, homogenous at all scales, 
with density
$\rho_{dark}$ (see fig.\ref{dm1}).
\begin{figure}
\epsfxsize 8cm
\centerline{\epsfbox{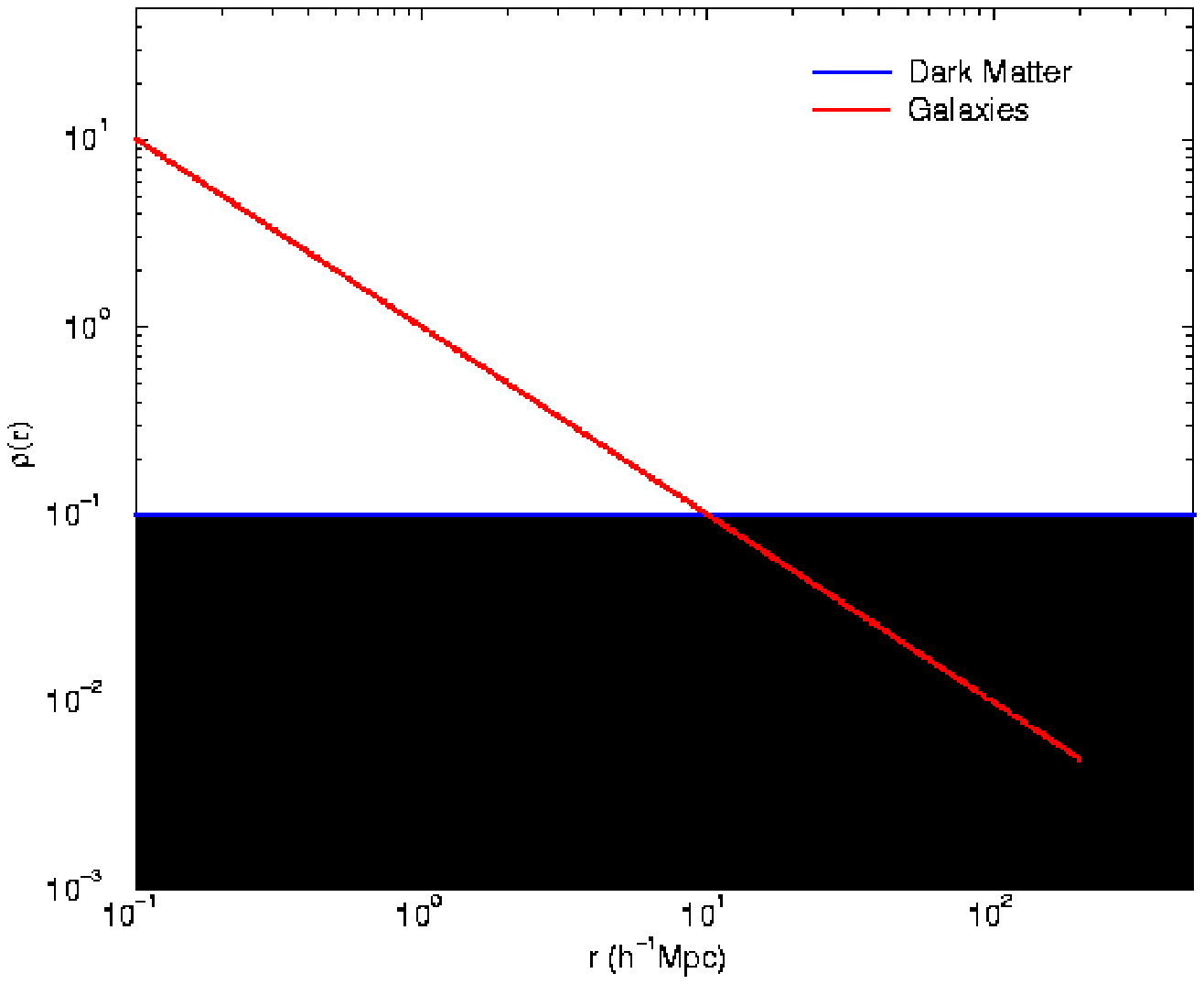}}
\caption{\label{dm1}   
 A simple two-component model: galaxies have a fractal distribution
corresponding to  correlated density fluctuations
superimposed to a uniform background. In the figure it is shown 
the behaviour of the number density of galaxies and the matter
density (from Durrer \& Sylos Labini 1998)} 
\eef 
For such a model there is a definite constant density at all scales larger
than $\lambda_0$.  This density $\rho_0$ is the sum of $\rho_{lum}$ and
$\rho_{dark}$.  This means that the behavior of this model at scales
larger than $\lambda_0$ is identical to that of the Friedmann model for which
\be
\label{yn1}
\Omega_0 = \Omega_{dark} + \Omega_{lum} \; .
\ee
At such large scales the Hubble law is unperturbed.  
The density distribution of luminous matter   for
scales $r < \lambda_0$, can be written as 
\be
\label{yn2}
\rho_{lum} (r) =   \rho_{lum} \left( \frac{\lambda_0 }{r}\right)^{\gamma}
\ee
where $\gamma=3-D$ as usual.
For the scales $r \ge \lambda_0$ we have that
\be
\label{yn3}
\rho_{lum} (r)  =  \rho_{lum}= \rho_{lum} (\lambda_0) \;.
\ee

The density contrast can be written as 
\be
\label{yn4}
\frac{\delta \rho  }{\rho_0} = \frac{ \rho_{lum} (r) + \rho_{dark} - \rho_0}
{\rho_0}
\ee
In terms of the Friedmann density parameters as defined above, this becomes,
at scales smaller than $\lambda_0$:
\be
\label{pek2}
\frac{\delta \rho }{  \rho_0} = 
\frac{\Omega_{lum} }{\Omega_0} 
\left(  
\left(\frac{r}{\lambda_0} \right)^{-\gamma}
- 1  \right) \; .
\ee
At scales larger than $\lambda_0$ the density contrast 
clearly vanishes. 

By using the linear approximation (Eq.\ref{pek1}), we may obtain
a rough estimation of the expected deflection from the Hubble law
in the two component model, for the scale
in which the density contrast is less than 1. 
Although the linear approximation
is valid only for $\delta \rho/\rho_0 \ltapprox 1$,
the obtained results give a first quantitative indication
of the effects of self-similar fluctuations.
Moreover the assumption of spherical 
mass distribution is a rough one, and it holds only 
for average quantities. In the case 
of real fractals deviation from spherical symmetry
can play an important role, at least at small scale.
Under these approximations,  
the radial velocity measured by an average observer
at scale $r < \lambda_0$ is (from Eq.\ref{pek1})
\be
\label{pek3}
V_{obs} = V_H \left(1-\left(\frac{1}{3}\right)
{\Omega_0}^{-0.4} 
 \Omega_{lum}
 \left( 
 \left( \frac{r}{\lambda_0} \right)^{-\gamma}
 - 1 \right)\right) 
\ee
Actually, this is the prediction averaged over many observers in different
fractal structure points (galaxies).  For any particular observer, there
will be a deflection from this average law.

We take the maximum scale $\lambda_0$ of
fractality and the fractal dimension $D$ from the observed
cosmological de Vaucouleurs Law and calculate the expected deflections
from the Hubble law in our two-components Friedmann model.  
In Fig.\ref{dm2} 
\begin{figure}
\epsfxsize 8cm
\centerline{\epsfbox{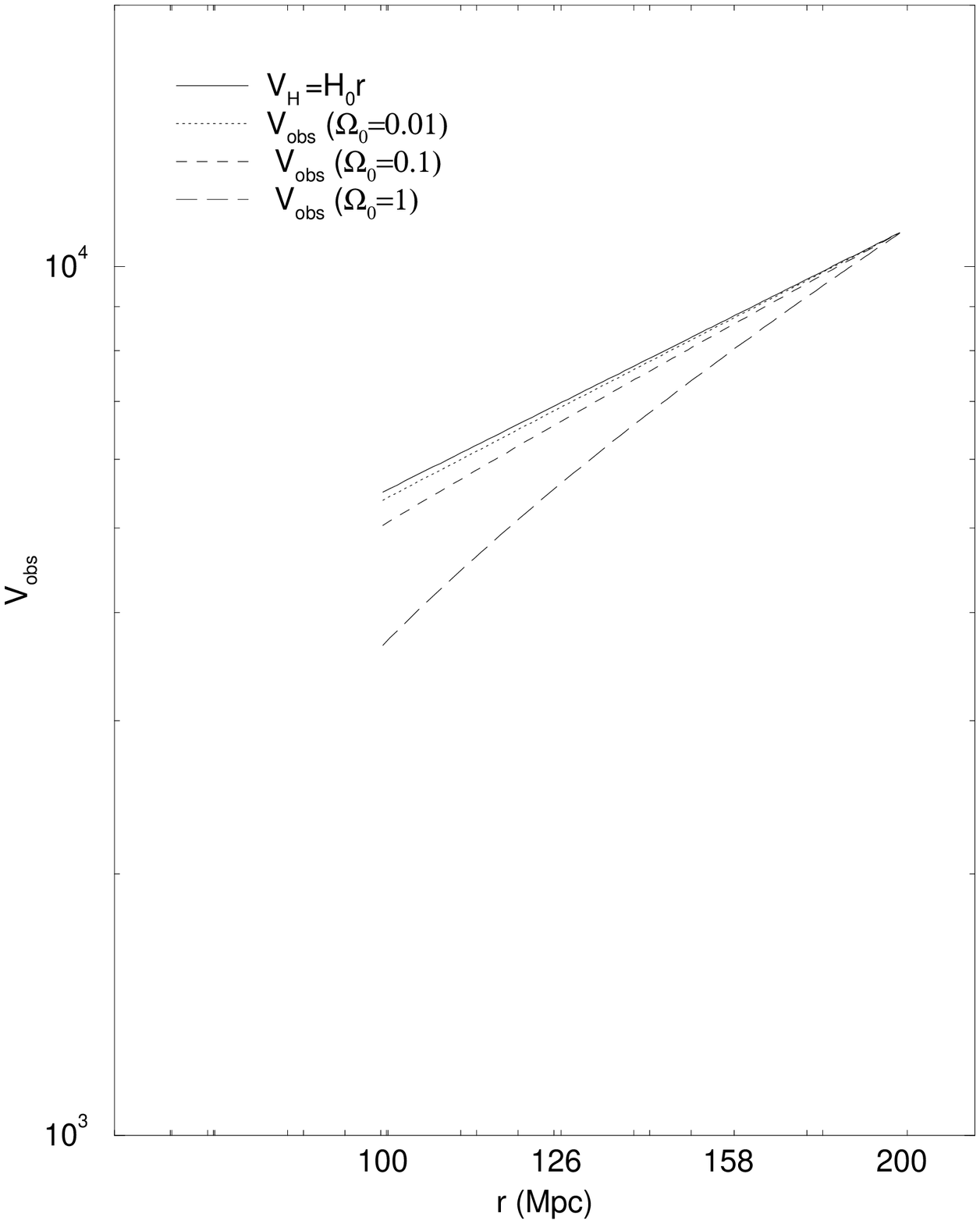}}
\caption{\label{dm2}  
We show three theoretical predictions for the velocity
deflection in the case where
the observed fractal structure contains all the matter, i.e. when
$\Omega_{lum} = \Omega_0$.  We have fixed $\lambda_0  = 200 Mpc$ and
fractal dimension $D = 2$.  The predictions correspond to three values
of the cosmological density parameter $\Omega_0 = 1, 0.1, 0.01$. 
(From Baryshev \etal, 1998) } 
\eef 
we show three theoretical predictions for the velocity
deflection in the case where
the observed fractal structure contains all the matter, i.e. when
$\Omega_{lum} = \Omega_0$.  We have fixed $\lambda_0 = 200 Mpc$ and
fractal dimension $D = 2$. 
In this case the linear approximation holds for
$r \gtapprox  130 Mpc$. At smaller scales we should
consider non-linear effects which are not simple to be treated.
However we should expect even stronger velocity
perturbations, due to the highly inhomogeneous 
structures distribution.
The predictions correspond to three values
of the cosmological density parameter $\Omega_0 = 1, 0.1, 0.01$.  From
Fig.\ref{dm2}
 it follows that such Friedmann models, purely fractal within
$200 Mpc$, are excluded if $\Omega_0 \gtapprox 0.01$.  This confirms the
previous suggestions that small $\Omega_0$ is needed for hierarchic
models 
(Sandage, \etal 1972 - hereafter STH).

For instance, there is one 
possible way to save the Friedmann universe with the
critical density parameter $\Omega_0 = 1$.  It was implied
already by STH that dark matter, uniformly filling the
whole universe and decreasing the relative density fluctuations, could
reconcile the observed fractal structure with the linear Hubble law.  
However, they did not give a quantitative estimate of the amount of dark
matter needed.  With
the new data on the Hubble and de Vaucouleurs laws, we can
derive the lower limit for the amount of the needed uniform dark matter.
In Eq.\ref{pek3} we fix 
$\Omega_0 = 1$ and
let $\Omega_{lum}$ have different values.  Fig.\ref{dm3}
\begin{figure}
\epsfxsize 8cm
\centerline{\epsfbox{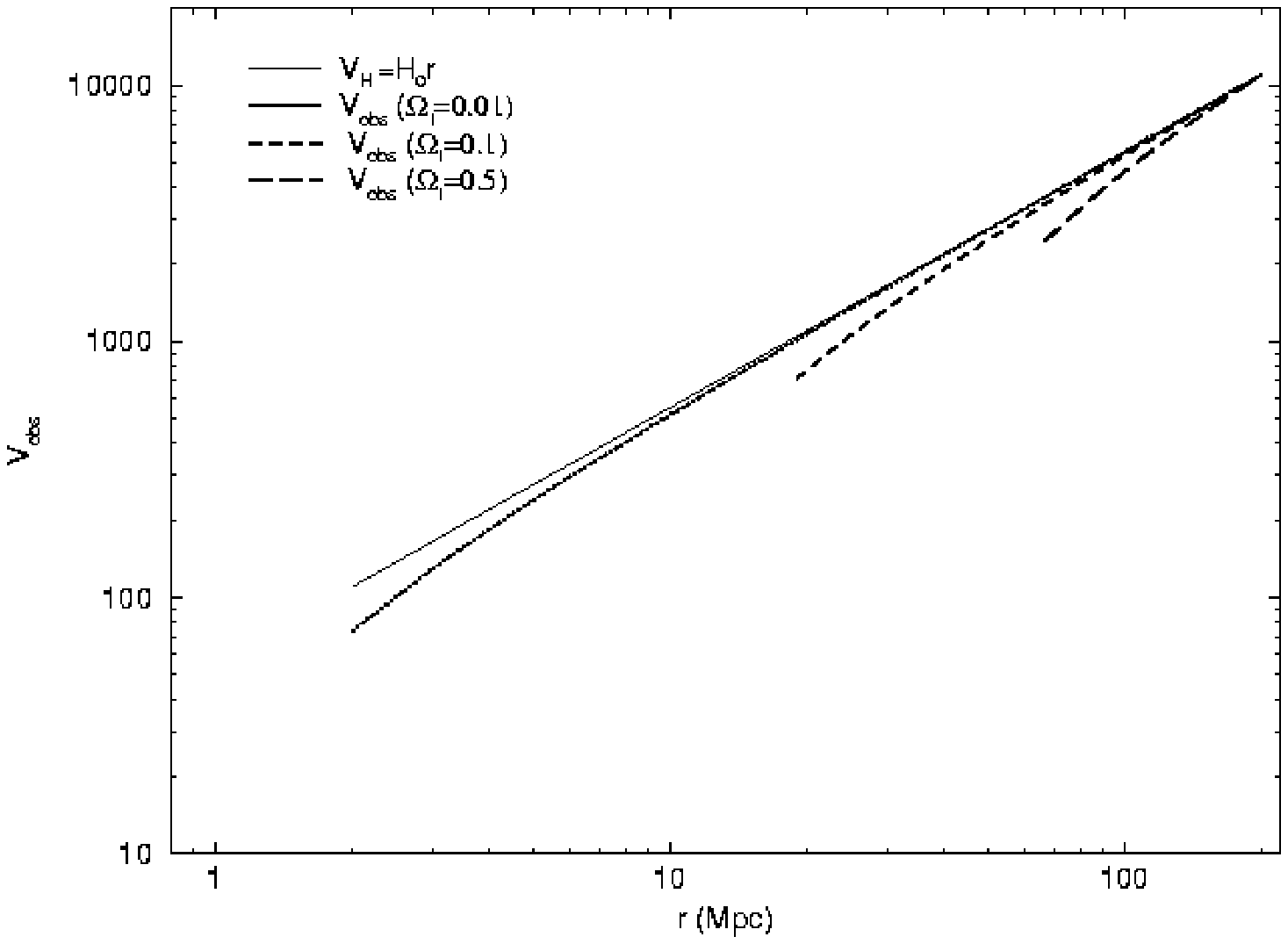}}
\caption{\label{dm3}  
In this case the total density is $\Omega_0=1$.
We show the behaviour of $V_{obs}$ derived by the 
linear perturbation  approximation in the case 
$\delta \rho / \rho_0 \ll 1$, for various values of $\Omega_{lum}$.
The fractal dimension of luminous matter is $D=$ up to
$\lambda_0 = 200 Mpc$. (From Baryshev \etal, 1998)} 
\eef 
gives the developed
version of the STH test, now showing that $\Omega_{dark}$ 
should be  larger than
0.99.  If the actual maximum scale of fractality is larger than $200 Mpc$
(with $D=2$),
then the amount of luminous matter may be in conflict with the Big Bang
nucleosynthesis prediction for baryonic matter.  For example, if
 $\lambda_0 \gtapprox 1000 Mpc$ 
(as suggested by Sylos Labini \etal, 1998)
then
$\Omega_{lum}$ will be probably less than 0.001.

It should be emphasized that this estimate of the amount of dark matter
is independent on the physics of the early universe.  It also does not
depend on the determination of mass-to-luminosity ratio of galaxies.

\section{Inconsistency of CDM models} 
It is simple to see
that a fractal behavior of galaxy distribution
with dimension $D \approx 2$ up to, at least, $\sim  50 \hmp$
is not compatible with standard CDM models.
In fig.\ref{cdm}
\bef  
\epsfxsize 10cm 
\epsfysize 10cm 
\centerline{\epsfbox{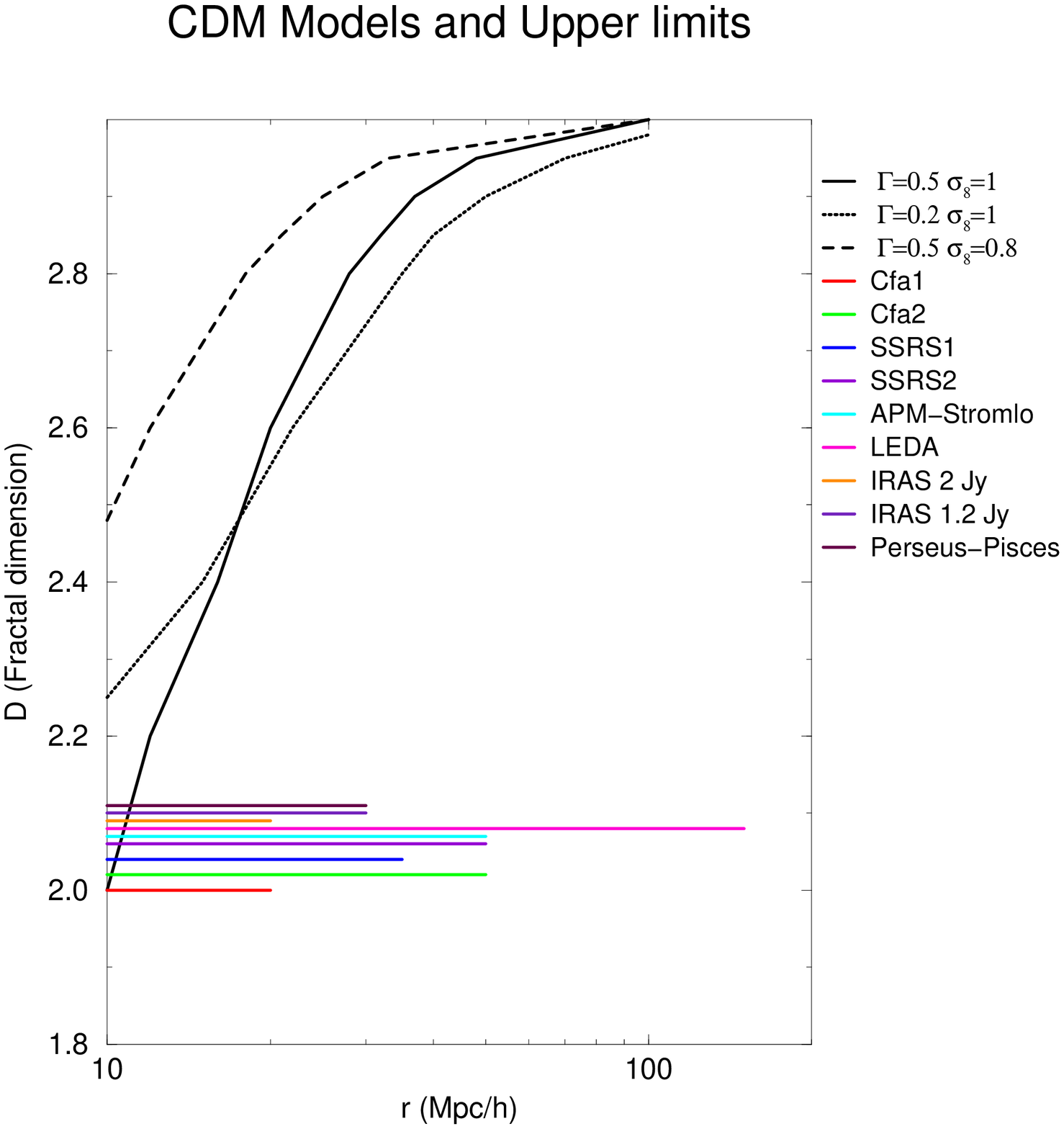}} 
\caption{\label{cdm} 
The fractal dimension versus distance in three
Cold Dark Matter models of power spectra 
with shape and normalized parameters as reported
in the labels (solid lines) (from Wu \etal, 1998).
We show also the different experimental determinations
of the fractal dimension we have obtained.
No agreement can be found at any scale. }
\eef
we show the behavior of the fractal dimension versus distance in three
Cold Dark Matter models of power spectra 
with shape and normalized parameters  (from Wu \etal, 1998).
We may see that a fractal dimension of 
$D \approx 2$ at $\sim 40 \div 50 \hmp$
is incompatible with all the models.
Probably by varying the parameters of
the simulation (or the mixture of Hot and Cold Dark Matter)
one may hope to obtain a better agreement.
Any new survey  has required a new adjustment
of the parameters  and
this alone shows the internal problems of the 
standard models of galaxy formation. 

We belive that the most important theoretical 
consequence of our results is that 
one may shift the attention
of the study from correlation amplitudes 
to  correlation exponents
(Sylos Labini \etal 1998,
and Durrer \& Sylos Labini, 1998).

\section{ Discussion}

Investigation of the large scale distribution of galaxies in 
the universe is now in a new phase, which is characterized 
by new observational data and new methods of analysis.  It has 
become an especially hot and debated topic in cosmology, because 
the revealed fractality contradicts Cosmological Principle 
in the sense of homogeneity but not in the sense of 
the equivalence of all the observers\cite{slmp98}. 
 Above we have discussed the fractality and its 
implications for cosmology.  Our main conclusions are: 
 
\begin{itemize}

\item Observations show that there is a fractal distribution of 
galaxies, having fractal dimension $D\approx 2$ in the scale 
range  from $1 Mpc \hmp $ to, at least, $100 \hmp$.  While there is a general
agreement on the small scale fractal properties of galaxy distribution,
the actual value of $D$ and 
the eventual presence of an upper cut-off, are still matter of 
debate\cite{slmp98} \cite{tee98}.  (See the web page
 {\it http://www.phys.uniroma1.it/DOCS/PIL/pil.html}
 where all these materials have been collected).

\item The traditional statistical analysis 
based on the assumption of homogeneity
(i.e. $\xi(r)$), 
should be replaced by the more general methods
of modern statistical physics.
Such methods are able to characterize scale-invariant
distributions
as well as regular ones.
 
\item An isotropic fractal distribution
is fully compatible with the reasonable
requirement of the equivalence of all the observers.
Hence the Standard Cosmological Principle,
which requires isotropy {\it and} homogeneity,
may be replaced by the  Conditional  Cosmological Principle.
In such a case the condition of local isotropy around
any structure point, without the assumption of 
analyticity of matter distribution, does not 
imply the homogeneity of matter distribution.

\item The paradox of linear Hubble law within the fractal de Vaucouleurs 
density-distance law is sharpened with the new data: strong deflections 
from the Hubble flow are expected in the framework of the standard 
Friedmann model. 
 
\item From a developed version of the old Sandage-Tammann-Hardy test we 
derive the minimum amount of the uniform dark matter, $\Omega_{dark} = 0.99$, 
which is consistent with the presently known Hubble and de Vaucouleurs 
laws.  This result is independent of the early universe physics.  If the 
maximum scale of fractality is larger than $100 \hmp$, this test may be 
regarded as crucial for the standard cosmology. 
 
\item We have shown
that a fractal behavior of galaxy distribution
with dimension $D \approx 2$ up to, at least, $\sim  50 \hmp$
is not compatible with standard CDM models.

\item One the most important theoretical 
consequence of our results is that 
one may shift the attention
of the study from correlation amplitudes 
to  correlation exponents.
The revision of the concept of bias (Durrer \& Sylos Labini, 1998)
is an example of such a situation. 
De Vega \etal \cite{sanchez1} \cite{sanchez2} have proposed a 
field theory approach to the fractal structure of the universe.
In such a model the dominant dynamical mechanism responsible
for the scale invariant distribution is self-gravity itself.
This model represents an interesting approach on the lines
of modern statistical physics previously:
it focuses on the expoenents rather than to the amplitudes
of correlation.
In the near future we are planning to 
study some experimental consequences of such a
theoretical framework.

\end{itemize}

\section*{Acknowledgemnts} 
I am in debt with Y. Baryshev, R. Durrer, M. Joyce, M. Montuori,
L. Pietronero and P. Teerikorpi 
 with whom various parts of this work have
been done. I warmly thank H. De Vega, H. Di Nella, A. Gabrielli,  
D. Pfenniger, 
N. Sanchez and F. Vernizzi for useful discussions and collaborations.
Finally I thank Prof. N. Sanchez and Prof. H. De Vega 
for their kind hospitality. This work has been partially supported by the 
EEC TMR Network  "Fractal structures and  self-organization"  
\mbox{ERBFMRXCT980183} and by the Swiss NSF.

\end {document}